\documentclass[prd, preprint, nofootinbib]{revtex4}

\usepackage{natbib}
\usepackage{graphicx}
\usepackage{amsfonts}
\usepackage{latexsym}
\usepackage{amsmath}
\usepackage{amssymb}
\usepackage{epsfig}

\newcommand{\bea}{\begin{eqnarray}}
\newcommand{\eea}{\end{eqnarray}}

\begin{document}

\title{ 
DAMPE excess from decaying right-handed neutrino dark matter
}

\author{Nobuchika Okada}
 \email{okadan@ua.edu}
 \affiliation{
Department of Physics and Astronomy, 
University of Alabama, Tuscaloosa, Alabama 35487, USA
}
\author{Osamu Seto}
 \email{seto@physics.umn.edu}
 \affiliation{Institute for International Collaboration, Hokkaido University, Sapporo 060-0815, Japan}
 \affiliation{Department of Physics, Hokkaido University, Sapporo 060-0810, Japan}
%

\begin{abstract}
The flux of high-energy cosmic-ray electrons plus positrons 
  recently measured by the DArk Matter Particle Explorer (DAMPE)
  exhibits a tentative peak excess at an energy of around $1.4$ TeV. 
In this paper, we consider the minimal gauged $U(1)_{B-L}$ model 
  with a right-handed neutrino (RHN) dark matter (DM) 
  and interpret the DAMPE peak with a late-time decay of the RHN DM into $e^\pm W^\mp$. 
We find that a DM lifetime $\tau_{DM} \sim 10^{28}$ s 
  can fit the DAMPE peak with a DM mass $m_{DM}=3$ TeV.  
This favored lifetime is close to the current bound on it by Fermi-LAT,
  our decaying RHN DM can be tested once the measurement of cosmic gamma ray flux is improved.  
The RHN DM communicates with the Standard Model particles 
  through the $U(1)_{B-L}$ gauge boson ($Z^\prime$ boson),  
  and its thermal relic abundance is controlled by only three free parameters: 
  $m_{DM}$, the $U(1)_{B-L}$ gauge coupling ($\alpha_{BL}$), and the $Z^\prime$ boson mass ($m_{Z^\prime}$). 
For $m_{DM}=3$ TeV, the rest of the parameters are restricted to be $m_{Z^\prime}\simeq 6$ TeV 
  and  $0.00807 \leq \alpha_{BL} \leq 0.0149$, 
  in order to reproduce the observed DM relic density and 
  to avoid the Landau pole for the running $\alpha_{BL}$ below the Planck scale. 
This allowed region will be tested by the search for a $Z^\prime$ boson resonance 
  at the future Large Hadron Collider. 
\end{abstract}


\preprint{EPHOU-17-016} 

\vspace*{2cm}
\maketitle


\section{Introduction}  
  
The nonvanishing neutrino masses have been established through various neutrino oscillation phenomena. 
The most attractive idea to explain the tiny neutrino masses is the so-called seesaw mechanism  
  with Majorana right-handed neutrinos (RHNs)~\cite{Seesaw}. 
The minimal gauged $U(1)_{B-L}$ model based on the gauge group 
   $SU(3)_C \times SU(2)_L \times U(1)_Y \times U(1)_{B-L}$~\cite{Mohapatra:1980qe} 
   is an elegant and simple extension of the Standard Model (SM), in which three RHNs are introduced 
  because of the gauge and gravitational anomaly cancellations
  and the Majorana masses of the RHNs are generated by the spontaneous breakdown 
  of the $U(1)_{B-L}$ gauge symmetry. 

The $B-L$ gauge symmetry breaking scale is yet undetermined and
 might be expected to be the TeV scale~\cite{Khalil:2006yi}. 
One theoretical ground which can naturally derive the TeV scale as symmetry breaking   
   is an extension of the minimal $U(1)_{B-L}$ model with the classically conformal invariance~\cite{Iso:2009ss}. 
In such a case, all new particles, the $U(1)_{B-L}$ gauge boson ($Z'$ boson), 
   the RHNs and the $B-L$ Higgs boson, have their masses of the order of TeV,  
   unless the $U(1)_{B-L}$ gauge coupling is extremely small.

Among three RHNs, one of those is a good candidate of dark matter (DM) if its lifetime
 is long enough compared to the age of the Universe because of tiny or vanishing Yukawa
 couplings~\cite{Khalil:2008kp, Anisimov:2008gg,Okada:2010wd,Kaneta:2016vkq,Biswas:2016bfo},
 while the remaining two RHNs need to have appropriate size of Yukawa couplings
 to generate active neutrino masses through the seesaw mechanism.
A RHN with tiny Yukawa couplings is a well-known typical example of decaying DM,
 for a recent review, see e.g., Refs.~\cite{Boyarsky:2012rt,Canetti:2012kh}.

Recently, the CALorimetric Electron Telescope (CALET) have reported new results
 of the flux of high-energy cosmic ray electrons plus positrons (CREs) measurement
 at the energy range from $10$ GeV to $3$ TeV~\cite{Adriani:2017efm}.
The DArk Matter Particle Explorer (DAMPE) have also reported new results
 in the similar energy range from $25$ GeV to $4.6$ TeV with
 very high-energy resolution~\cite{Ambrosi:2017wek}.
On the one hand, those are consistent with the results by 
  the AMS-02~\cite{Aguilar:2014mma} and Fermi-LAT~\cite{Abdollahi:2017nat} measurements 
  in a wide energy range.
On the other hand, interestingly,
 the observed spectrum by the DAMPE exhibits not only a spectral break around $0.9$ TeV,
 which was previously reported by the H.E.S.S.~\cite{Aharonian:2008aa}, 
 but also a tentative peak excess at an energy of around $1.4$ TeV.
Motivated by the DAMPE results, possible sources of the CREs excesses 
  have been investigated.  
Although supernova remnant or pulsar (wind nebula) may explain both the excess and the peak at $1.4$ TeV~\cite{Yuan:2017ysv,Fang:2017tvj}, the interpretation with DM annihilations or decays is another interesting possibility.  
In addition to examination of implications of the DAMPE data to DM properties~\cite{Zu:2017dzm,Athron:2017drj,Huang:2017egk,Jin:2017qcv,Gao:2017pym,Niu:2017hqe,Yang:2017cjm,Ge:2017tkd,ZhaoFangSuMiller} , 
  many specific particle physics models have been proposed~\cite{Fan:2017sor,Gu:2017gle,Duan:2017pkq,Tang:2017lfb,Chao:2017yjg,Gu:2017bdw,Cao:2017ydw,Duan:2017qwj,
Liu:2017rgs,Chao:2017emq,Chen:2017tva,Li:2017tmd,Gu:2017lir,Nomura:2017ohi,Zhu:2017tvk,
Ghorbani:2017cey,Cao:2017sju,Ding:2017jdr,Liu:2017obm}.

In this paper, we revisit the minimal gauged $U(1)_{B-L}$ model with a RHN DM~\cite{Okada:2010wd} 
 and show the DAMPE peak excess can be explained by the RHN DM decay.
Our model is distinctive from other models recently proposed to account for the DAMPE peak excess at $1.4$ TeV,
  because interactions between the RHN DM and the SM particles are not leptophilic.
The RHN DM interacts with all quarks and leptons 
  through the $Z^{\prime}$ boson and its relic abundance is obtained 
  via the usual thermal freeze-out process. 
The CREs are created by a late-time decay of the RHN DMs into $e^\pm W^\mp$ 
   through a tiny neutrino Dirac Yukawa coupling of the RHN DM.

This paper is organized as follows: 
In the next section, we first describe our model. 
Then, we calculate the thermal relic density of the RHN DM and 
  the $Z^\prime$ boson production cross-section at the LHC Run-2. 
By imposing the constraints from the observed DM relic density, 
  the LHC results of the search for a narrow resonance 
  and the perturbativity condition of the $U(1)_{B-L}$ gauge coupling 
  up to the Planck scale, we identify a parameter region of the model.
In Sec.~\ref{sec:3}, we consider the late-time RHN DM decay and 
  identify the DM lifetime and DM mass to fit the DAMPE peak. 
In the last section, we summarize our results and discuss an implication 
  of our scenario on the future LHC.

\section{The Minimal $B-L$ Model with a RHN DM}
\label{sec:2}

\subsection{The model}

\begin{table}[t]
\begin{center}
\begin{tabular}{|c|ccc|c|}
\hline
      &  SU(3)$_c$  & SU(2)$_L$ & U(1)$_Y$ & U(1)$_{B-L}$  \\ 
\hline
$q^{i}_{L}$ & {\bf 3 }    &  {\bf 2}         & $ 1/6$       & $1/3 $   \\
$u^{i}_{R}$ & {\bf 3 }    &  {\bf 1}         & $ 2/3$       & $1/3 $   \\
$d^{i}_{R}$ & {\bf 3 }    &  {\bf 1}         & $-1/3$       & $1/3 $  \\
\hline
$\ell^{i}_{L}$ & {\bf 1 }    &  {\bf 2}         & $-1/2$       & $-1 $    \\
$e^{i}_{R}$    & {\bf 1 }    &  {\bf 1}         & $-1$         & $-1 $   \\
\hline
$H$            & {\bf 1 }    &  {\bf 2}         & $- 1/2$       & $0 $   \\  
\hline
$N^{i}_{R}$    & {\bf 1 }    &  {\bf 1}         &$0$                    & $-1$     \\
$\Phi$            & {\bf 1 }       &  {\bf 1}       &$ 0$                  & $ + 2 $  \\ 
\hline
\end{tabular}
\end{center}
\caption{
The particle content of the minimal $U(1)_{B-L}$ model. 
In addition to the SM particle content ($i=1,2,3$), the three right-handed neutrinos  
  ($N_R^i$ ($i=1, 2, 3$)) and the $U(1)_{B-L}$ Higgs field ($\Phi$) are introduced.   
}
\label{table1}
\end{table}

Our model is based on the gauge group $SU(3)_C \times SU(2)_L \times U(1)_Y \times U(1)_{B-L}$, and
 the three right-handed neutrinos ($N_R^i$) and one SM singlet $B-L$ Higgs field ($\Phi$) are introduced.
The particle content is listed in Table~\ref{table1}.
Due to the additional gauge symmetry $U(1)_{B-L}$, 
 the covariant derivative for each field is given by 
\begin{equation}
 D_{\mu}= D_{\mu}^{(SM)} -i q_{B-L}g_{B-L}Z^\prime_{\mu},
\end{equation}
 where $D_{\mu}^{(SM)}$ is the covariant derivative for the SM gauge group, 
 and $q_{B-L}$ is the charge of each field under the $U(1)_{B-L}$
 with its gauge coupling $g_{B-L}$.
The Yukawa sector of the SM is extended to have 
\bea
\mathcal{L}_{Yukawa} \supset  - \sum_{i=1}^{3} \sum_{j=1}^{3} Y^{ij}_{D} \overline{\ell^i_{L}} H N_R^j 
          -\frac{1}{2} \sum_{k=1}^{3} Y_{N^k} \Phi \overline{N_R^{k~C}} N_R^k 
           + {\rm H.c.} ,
\label{Lag1} 
\eea
where the first term is the neutrino Dirac Yukawa coupling, and the second is the Majorana Yukawa coupling. 
Without loss of generality, the Majorana Yukawa couplings are already diagonalized in our basis.  
Once the $U(1)_{B-L}$ Higgs field $\Phi$ develops a nonzero vacuum expectation value,
 the $U(1)_{B-L}$ gauge symmetry is broken and the Majorana mass terms
 of the RHNs are generated. 
Then, the seesaw mechanism is automatically implemented in the model after the electroweak symmetry breaking.     
Let us assume that $Y_D^{13}$ is extremely small while $Y_D^{23} =Y_D^{33} = 0$, 
  in which case the RHN $N_R^3$ becomes quasi-stable and plays the role of dark matter~\cite{Anisimov:2008gg}. 
Note that a $Z_2$-parity emerges in the limit of $Y_D^{13} \to 0$~\cite{Okada:2010wd} 
  and $N_R^3$ becomes stable. 
Since $N_R^3$ has a negligible contribution to the seesaw mechanism with $Y_D^{13} \ll 1$, 
 only two RHNs ($N_R^{1,2}$) account for the neutrino mass generation via the seesaw mechanism. 
This system is approximately the so-called minimal seesaw~\cite{King:1999mb, Frampton:2002qc}, 
 which is the minimal setup to reproduce the observed neutrino oscillation data 
 with a prediction of one massless neutrino. 
Hereafter,  we call the RHN DM ($N_R^3$) as ``$\chi$''. 

The renormalizable scalar potential for $H$ and $\Phi$ is given by 
\bea  
V = \lambda_H \left(  H^{\dagger}H -\frac{v^2}{2} \right)^2
+ \lambda_{\Phi} \left(  \Phi^{\dagger} \Phi -\frac{v_{BL}^2}{2}  \right)^2
+ \lambda_{\rm mix} 
\left(  H^{\dagger}H -\frac{v^2}{2} \right) 
\left(  \Phi^{\dagger} \Phi -\frac{v_{BL}^2}{2}  \right) , 
\label{Higgs_Potential}
\eea
where all quartic couplings are chosen to be positive for simplicity. 
At the potential minimum, the Higgs fields develop their vacuum expectation values as 
\bea
  \langle H \rangle =  \left(  \begin{array}{c}  
    \frac{v}{\sqrt{2}} \\
    0 \end{array}
\right),  \;  \;  \; \; 
\langle \Phi \rangle =  \frac{v_{BL}}{\sqrt{2}} ,
\eea
 with $v \simeq 246$ GeV.
Associated with the U(1)$_{B-L}$ symmetry breaking, the Majorana neutrinos $N_R^j$ $(j=1,2)$, 
  the DM particle $\chi$, the $Z^\prime$ gauge boson, respectively, acquire their masses as 
\bea 
  m_{N_R^{1,2}}=\frac{Y_{N_R^{1,2}}}{\sqrt{2}} v_{BL},  \; \; 
  m_{\chi}=\frac{Y_{N_R^3}}{\sqrt{2}} v_{BL},  \; \; 
  m_{Z^\prime} = 2 g_{BL} v_{BL}.  
\eea  
In the vacuum,
 the SM-like Higgs boson $h$ and the $B-L$ Higgs-like scalar $\phi$ also have their masses as 
\begin{eqnarray}
m_h^2 &=& 2\lambda_H v^2 \cos^2\alpha + 2\lambda_{\Phi} v_{BL}^2 \sin^2\alpha 
          - 2\lambda_\mathrm{mix} v v_{BL} \sin\alpha\cos\alpha , \label{Mh} \\
m_{\phi}^2 &=& 2\lambda_H v^2 \sin^2\alpha + 2\lambda_{\Phi} v_{BL}^2 \cos^2\alpha
          + 2\lambda_\mathrm{mix} v v_{BL} \sin\alpha\cos\alpha . \label{MH}
\end{eqnarray}
 with a mixing angle $\alpha$ given by 
\begin{equation}
 \tan2\alpha
    = - \frac{\lambda_\mathrm{mix} v v_{BL}}{(\lambda_H v^2 -\lambda_{\Phi} v_{BL}^2 )} .
\end{equation}
We fix $m_h \simeq 125$ GeV.
Because of the LEP constraint $v_{BL} \gtrsim 6$ TeV \cite{Carena:2004xs, Heeck:2014zfa}, 
  the natural scale of $m_{\phi}$ and $m_{\chi}$ is of the order of TeV.
We consider the mass spectrum $m_{\phi} > m_{\chi} =\mathcal{O}(1)$ TeV.
In addition, the mass mixing of the $Z^\prime$ boson with the SM $Z$ boson
  is very small, hence we neglect it in our analysis.  

RHNs have tiny mixings with left-handed components and,
 through those mixings, RHNs decay to SM particles by the weak interaction. 
The mixing angles are given by the ratio of the Dirac masses to the Majorana masses.
The mixing angle of $\chi$ is expressed as
\begin{equation}
\mathcal{R} = \frac{Y_D^{13}v}{Y_{N_R^3}v_{BL}}.
\end{equation}
$\chi$ has the charged current, the neutral current, and 
  the Yukawa interactions are as follows: 
\bea 
\mathcal{L}_{int} & \supset& 
 -\frac{g}{\sqrt{2}} W_{\mu}^+ \; \overline{e} \gamma^{\mu} P_L \mathcal{R} \chi
 - \frac{g}{2 \cos \theta_{\rm W}}  Z_{\mu}  \; \overline{\nu_e} \gamma^{\mu} P_L \mathcal{R} \chi
 - \frac{Y^{13}_D}{\sqrt{2}} h \; \overline{\nu_e} P_L \mathcal{R} \chi , 
\label{Nint}  
\eea
where $e$ and $\nu_e$ are the electron and electron type neutrinos, 
  $P_L =  (1- \gamma_5)/2$,  and $\theta_{\rm W}$ is the weak mixing angle. 
Through the above interactions, $\chi$ decays 
  into $e W$, $\nu_e Z$, and $\nu_e h$ with the corresponding partial decay widths: 
\bea
\Gamma(\chi \rightarrow e W)
 &=& \frac{\mathcal{R}^2}{16 \pi} 
 \frac{ (m_{\chi}^2 - m_W^2)^2 (m_{\chi}^2+2 m_W^2)}{m_{\chi}^3 v^2} , 
\nonumber \\
\Gamma(\chi \rightarrow \nu_e Z)
 &=& \frac{\mathcal{R}^2}{32 \pi} 
 \frac{ (m_{\chi}^2 - m_Z^2)^2 (m_{\chi}^2+2 m_Z^2)}{m_{\chi}^3 v^2} ,
\nonumber \\
\Gamma(\chi \rightarrow \nu_e h)
 &=& \frac{\mathcal{R}^2}{32 \pi}\frac{(m_{\chi}^2-m_h^2)^2}{m_{\chi} v^2} . 
\label{widths}
\eea 

In this paper, we take $\lambda_{\rm mix} \ll1$ in Eq.~(\ref{Higgs_Potential}), 
  so that the RHN DM communicates with the SM particles 
  mainly through the $Z^\prime$ boson.
In this ``$Z^\prime$-portal RHN DM'' scenario \cite{OO}, 
  only three free parameters ($\alpha_{BL}=g_{BL}^2/(4 \pi)$, $m_{Z^\prime}$, and $m_{\chi}$) 
  are involved in our analysis. 
As we will discuss in the following, it turns out that the RHN DM mass must be very close 
   to one half of the $Z^\prime$ boson mass ($m_{\chi} \simeq m_{Z^\prime}/2$)   
   in order to reproduce the observed DM relic density.
Thus, our results are effectively described by only two free parameters: $\alpha_{BL}$ and $m_{Z^\prime}$.

\subsection{Cosmological constraints on $Z^\prime$-portal RHN DM}

In this subsection, we evaluate the DM thermal relic density and 
  identify an allowed parameter region to obtain it 
  in the range of $0.1183 \leq \Omega_{DM} h^2 \leq 0.1213$ (68\% confidence level), 
  measured by the Planck satellite experiment~\cite{Ade:2015xua}.
The DM relic abundance is evaluated by integrating the Boltzmann equation given by \cite{KolbTurner}
\bea 
  \frac{dY}{dx}
  = - \frac{x s \langle \sigma v_{\rm rel} \rangle}{H(m_{DM})} \left( Y^2-Y_{EQ}^2 \right), 
\label{Boltmann}
\eea  
where $m_{DM}$ denotes the mass of DM particle,
 $x=m_{DM}/T$ is the ratio of the DM mass to the temperature of the Universe ($T$), 
 $H(m_{DM})$ is the Hubble parameter at $T=m_{DM}$, 
 $Y$ is the yield (the ratio of the DM number density to the entropy density $s$) of 
 the DM, and $Y_{EQ}$ is the yield of the DM particle in thermal equilibrium. 
The thermal average of the DM annihilation cross-section times relative velocity, 
  $\langle \sigma v_{\rm rel} \rangle$,  is given by 
\bea 
\langle \sigma v_{\rm rel} \rangle = \left(n_{EQ} \right)^{-2} g_{DM}^2
  \frac{m_{DM}}{32 \pi^4 x} 
  \int_{4 m_{DM}^2}^\infty  ds \; \sigma(s) \,  \sqrt{1- \frac{4 m_{DM}^2}{s}} \, s^{3/2} \, K_1 \left(\frac{x \sqrt{s}}{m_{DM}}\right) , 
\label{ThAvgSigma}
\eea
where $n_{EQ}$ is the DM  number density in thermal equilibrium, 
  $g_{DM}=2$ is the internal degrees of freedom for the DM, and  
  $K_1$ is the modified Bessel function of the first kind. 

In our model, assuming $m_{N_R^{1,2}} > m_{\chi},$ $m_{Z^\prime}/2$,  for simplicity, 
  the total cross-section $\sigma(s)$ of all relevant DM pair annihilation modes
 $\chi \chi  \to Z^\prime \to f {\bar f}$ 
 ($f$ denotes the SM fermions) is calculated as 
\bea 
 \sigma(s)=\frac{13 \pi}{3}  \alpha_{BL}^2  \frac{\sqrt{s (s-4 m_{\chi}^2)}}
  {(s-m_{Z^\prime}^2)^2+m_{Z^\prime}^2 \Gamma_{Z^\prime}^2} .
\label{DMSigma}
\eea  
Here, we note that other channels such as $\chi \chi  \to Z^\prime \to Z' \phi$
 or $\chi \chi  \to Z^\prime \to \phi \phi$ are kinematically forbidden
 and the total decay width of $Z^\prime$ boson is given by
\bea
\Gamma_{Z'} = 
 \frac{\alpha_{BL}}{6} m_{Z^\prime} 
 \left[ 13 +  \left( 1- \frac{4 m_{\chi}^2}{m_{Z^\prime}^2} \right)^{\frac{3}{2}} 
 \theta \left( \frac{m_{Z^\prime}}{m_\chi} - 2 \right)  \right] ,
\label{width}
\eea
  where $\theta(x)$ is the unit step function, and the masses of all SM fermions are neglected. 
 
\begin{figure}[t]
\begin{center}
\includegraphics[width=0.6\textwidth,angle=0,scale=1.06]{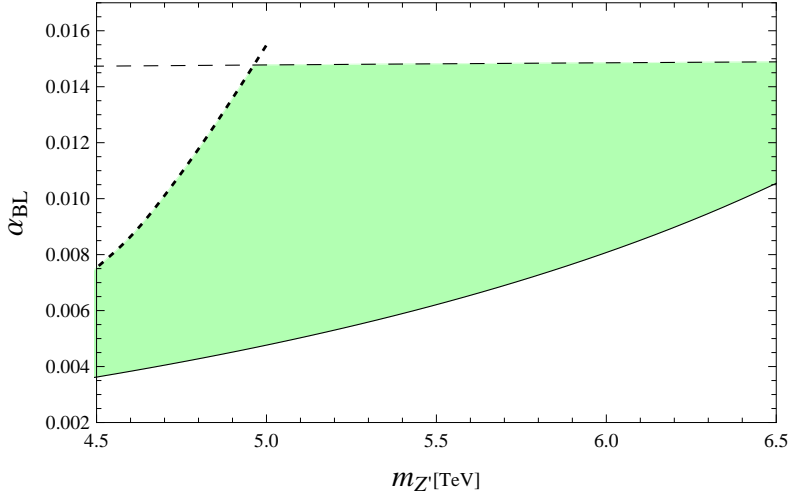} 
\end{center}
\caption{
Allowed parameter region (shaded) for the $Z^\prime$-portal RHN DM scenario. 
The (black) solid line denotes the lower bound on $\alpha_{BL}$ as a function of $m_{Z^\prime}$ 
    to reproduce the observed  DM relic density. 
The diagonal dotted line shows the upper bound on $\alpha_{BL}$ 
   obtained from the ATLAS results on the search for a narrow resonance 
   with the combined dielectron and dimuon final states at the LHC Run-2. 
The perturbativity bound on $\alpha_{BL}$ is depicted as the horizontal dashed line. 
}
\label{fig:1}
\end{figure}

Now, we solve the Boltzmann equation numerically, and 
   find the asymptotic value of the yield $Y(\infty)$ to evaluate the present DM relic density as 
\bea 
  \Omega_{DM} h^2 =\frac{m_{DM} s_0 Y(\infty)} {\rho_c/h^2}, 
\eea 
  where $s_0 = 2890$ cm$^{-3}$ is the entropy density of the present Universe, 
  and $\rho_c/h^2 =1.05 \times 10^{-5}$ GeV/cm$^3$ is the critical density. 
Three free parameters, $\alpha_{BL}$, $m_{Z^\prime}$, and $m_{\chi}$, are involved in our analysis. 
For $m_{\chi}$ at the TeV scale, we find that the enhancement of the DM annihilation cross-section 
  via the $Z^\prime$ boson resonance is necessary to reproduce the observed DM density. 
Hence, we always find $m_{\chi} \simeq m_{Z^\prime}/2$. 
In Fig.~\ref{fig:1} we show a parameter region to reproduce the observed DM relic density 
  in the range of $0.1183 \leq \Omega_{\chi} h^2 \leq 0.1213$. 
The solid line represents the lower bounds on $\alpha_{BL}$ as a function of $m_{Z^\prime}$.

\subsection{Constraints from $Z^\prime$ boson search at the LHC Run-2}
\label{sec:5}

The ATLAS and CMS collaborations have been searching for a narrow resonance 
 with dilepton final states at the LHC Run-2. 
In their analysis, the so-called sequential SM $Z^\prime$ ($Z^\prime_{SSM}$) 
 has been studied as a reference model, 
 assuming the $Z^\prime_{SSM}$ boson has exactly the same properties 
 as the SM $Z$ boson, except for its mass.  
In the following, we interpret the current LHC constraints on the $Z^\prime_{SSM}$ boson 
 into the $U(1)_{B-L}$ $Z^\prime$ boson to identify an allowed parameter region.  
We employ the latest upper bound on the $Z^\prime_{SSM}$ production cross-section 
 reported by the ATLAS collaboration~\cite{ATLAS:2017}.

The cross-section for the process $pp \to Z^\prime +X \to \ell^{+} \ell^{-} +X$  is given by
\begin{eqnarray}
 \sigma
 =  \sum_{q, {\bar q}}
 \int d M_{\ell \ell} 
 \int^1_ \frac{M_{\ell \ell}^2}{s} dx
 \frac{2 M_{\ell \ell}}{x s}  
 f_q(x, Q^2) f_{\bar q} \left( \frac{M_{\ell \ell}^2}{x s}, Q^2
 \right)  {\hat \sigma} (q \bar{q} \to Z^\prime \to  \ell^+ \ell^-) ,
\label{X_LHC}
\end{eqnarray}
where $M_{\ell \ell}$ is the invariant mass of a final state dilepton,  
  $f_q$ is the parton distribution function for a parton (quark) ``$q$'', 
  and $\sqrt{s} =13$ TeV is the center-of-mass energy of the LHC Run-2.
In our numerical analysis, we employ CTEQ6L~\cite{CTEQ} 
  for the parton distribution functions with the factorization scale $Q= m_{Z^\prime}$. 
The cross-section for the colliding partons is given by
\bea 
{\hat \sigma}(q \bar{q} \to Z^\prime \to  \ell^+ \ell^-) =
 \frac{4 \pi}{81}  \alpha_{BL}^2 
 \frac{M_{\ell \ell}^2}{(M_{\ell \ell}^2-m_{Z^\prime}^2)^2 + m_{Z^\prime}^2 \Gamma_{Z^\prime}^2} , 
\label{X_LHC2}
\eea
where ``$q$'' being the up-type ($u$) and down-type ($d$) quarks, respectively.

In interpreting the latest ATLAS results \cite{ATLAS:2017} 
  on the $Z^\prime_{SSM}$ boson into the U(1)$_{B-L}$ $Z^\prime$ boson case, 
  we follow the strategy in Ref.~\cite{OO}: 
  we first calculate the cross-section of the process $pp \to Z^\prime_{SSM} +X \to \ell^{+} \ell^{-} +X$ 
  by Eq.~(\ref{X_LHC}) and then we scale our result by a $k$-factor ($k = 1.28$) 
  so as to match our cross-section result with the theoretical prediction 
  of the cross-section presented in the ATLAS paper \cite{ATLAS:2017}. 
With the $k$-factor determined in this way, we calculate the cross-section 
   for the process $pp \to Z^\prime+X \to \ell^{+} \ell^{-} +X$  
   to identify an allowed region for the model parameters: $\alpha_{BL}$ and $m_{Z^\prime}$.

As a theoretical constraint, we may impose an upper bound on the U(1)$_{B-L}$ gauge coupling 
   to avoid the Landau pole in its renormalization group evolution $\alpha_{BL}(\mu)$ 
  up to the Plank scale of $M_{Pl}=1.22 \times 10^{19}$ GeV, 
  namely, $1/\alpha_{BL}(M_{Pl}) > 0$.  
We define the gauge coupling $\alpha_{BL}$ used in our analysis 
  for the dark matter physics and LHC physics 
  as the running gauge coupling $\alpha_{BL}(\mu)$ at $\mu=m_{Z^\prime}$. 
Employing the renormalization group equation at the one-loop level 
  with $m_{N_R^1}=m_{N_R^2}=m_\phi=m_{Z^\prime}$, for simplicity, we find 
\bea
  \alpha_{BL} < \frac{\pi}{6 \ln \left[ \frac{M_{Pl}}{m_{Z^\prime}} \right]}. 
  \label{pert} 
\eea 
In Fig.~\ref{fig:1}, the diagonal dotted line is the upper bound on $\alpha_{B-L}$ 
 obtained from the ATLAS results on the search for a narrow resonance 
 with the combined dielectron and dimuon final states at the LHC Run-2. 
The perturbativity bound of Eq.~(\ref{pert}) is depicted as the horizontal dashed line. 
The shaded region is the available region consistent
 with the observed DM density, the LHC constraints, and the perturbativity bound.

\section{Decaying Dark Matter Interpretation}
\label{sec:3}

The CREs propagation 
 is described by the following diffusion equation:
\begin{equation}
\frac{\partial}{\partial t} f - K(E)\nabla^2 f -\frac{\partial}{\partial E}\left( b(E) f \right) 
 = Q(E, \mathbf{x}) ,
\end{equation}
 where $f = d N/d E$ is the energy spectrum of electrons and positrons,
 $K(E)$ is the diffusion coefficient and
 $b(E)$ is the energy loss rate~\cite{Baltz:1998xv}.
When we consider a DM annihilation or a decay as a source of the CREs, 
 the source term is given by 
\begin{equation}
 Q(E, \mathbf{x}) = \frac{1}{2}\frac{\rho_{DM}^2(\mathbf{x})}{m_{DM}^2}(\sigma v_{\rm rel})_0 \, 
 {\rm Br}^\mathrm{ann} \, \frac{dN}{dE} ,
\label{Eq:source:ann}
\end{equation} 
 for the DM annihilation and  
\begin{equation}
 Q(E, \mathbf{x}) = \frac{\rho_{DM}(\mathbf{x})}{m_{DM}} \frac{1}{\tau_{DM}}  \,
   {\rm Br}^\mathrm{dec} \, \frac{dN}{dE} ,
\label{Eq:source:dec}
\end{equation}
for the DM decay, respectively.
Here, $(\sigma v_{\rm rel})_0$ is the DM annihilation cross-section today, which is the s-wave limit or equivanetly the zero-temperature limit of the thermal averaged cross-section $\langle\sigma v_{\rm rel}\rangle$, $\tau_{DM}$ is the DM lifetime, 
 and the branching fractions into electron/positron final states 
 from the DM annihilation and decay are denoted 
 as ${\rm Br}^\mathrm{ann}$ and ${\rm Br}^\mathrm{dec}$, respectively.

Since electrons and positrons cannot travel a long distance 
  without dissipation of energy, we need to assume a nearby strong source 
  for the DM interpretation of the peak in the DAMPE data. 
We consider that the CREs originate from
  a nearby local sub-halo a.k.a local clump, apart from us about $0.1$ kpc, 
  as commonly assumed in the proposal of the interpretation for the peak by DM annihilations. 
Using Eqs.~(\ref{Eq:source:ann}) and (\ref{Eq:source:dec}) with
 ${\rm Br}^\mathrm{ann} \sim {\rm Br}^\mathrm{dec} \sim 1$, 
  we roughly estimate the DM lifetime to generate the same intensity of the peak 
  from the DM annihilation: 
\begin{align}
\tau_{DM} & \sim  \left(\frac{\overline{\rho_{DM}}}{m_{DM}}\frac{(\sigma v_{\rm rel})_0}{2} \right)^{-1} \nonumber \\
  & \sim  2\times 10^{28} \,\mathrm{sec} \; 
 \left( \frac{10  \, \mathrm{GeV/cm^3}}{\overline{\rho_{DM}}} \right)
 \left(\frac{m_{DM}}{3 \, \mathrm{TeV}}\right)\left(\frac{3\times 10^{-26} \, \mathrm{cm{}^3/s}}{(\sigma v_{\rm rel})_0}\right). 
\label{eq:lifetime}
\end{align}
where $\overline{\rho_{DM}}$ is given by  
\begin{align}
\overline{\rho_{DM}} = \frac{\int dV \; \rho_{DM}^2}{\int dV \; \rho_{DM}}.
\end{align}
Here, we have taken $m_{\chi}=3$ TeV for the RHN DM mass 
  to produce a monochromatic electron/position with an energy 1.5 TeV by its decay $\chi \to e^\pm W^\mp$. 
According to the results in Ref.~\cite{Yuan:2017ysv},
 the required density $\overline{\rho_{DM}}$ is about $17$-$35$ times 
   of the canonical local density of $\rho_0=0.4 \, \mathrm{GeV/cm^3}$  
   to give the DAMPE peak from the DM annihilation 
   with a typical thermal cross-section, $\langle\sigma v_{\rm rel}\rangle = 3\times 10^{-26} \, \mathrm{cm{}^3/s}$.

With the mass of $m_{DM}=3$ TeV, the RHN DM has three decay modes 
  $\chi \to e^\pm W^\mp$, $\nu_e Z$ and $\nu_e h$ with the branching ratios 
  of $50\%$, $25\%$ and $25\%$, respectively. 
This DM decay can also produce diffused gamma rays from hadrons generated 
  by the cascade decay of  the weak bosons ($W^\pm$ and $Z$) and the Higgs boson ($h$). 
Gamma ray flux from dwarf spheroidal galaxies (dSphs)~\cite{Baring:2015sza}
 as well as galactic and extragalactic~\cite{Cohen:2016uyg} provide us with
 the lower limit on the DM lifetime.    
We refer the results in Ref.~\cite{Cohen:2016uyg} and find the lower limit on the RHN DM lifetime as   
\begin{equation}
 \tau_{DM} > \mathrm{a \; few} \times 10^{28} \,\mathrm{s},
\label{lifetime:constr}
\end{equation}
 for a few TeV mass of DM.
The favored value of lifetime Eq.~(\ref{eq:lifetime}) looks marginally consistent with this constraint. 
Furthermore, neutrino flux with a peak at around $1.5$ TeV also is generated through the decay. 
The strength is completely fixed by the above decaying branching ratio.
Taking into account various uncertainties such as the intensity of CREs flux or DM density,
 the DAMPE peak can be explained by our decaying RHN DM.

\begin{figure}[t]
\begin{center}
\includegraphics[width=0.6\textwidth,angle=0,scale=1.06]{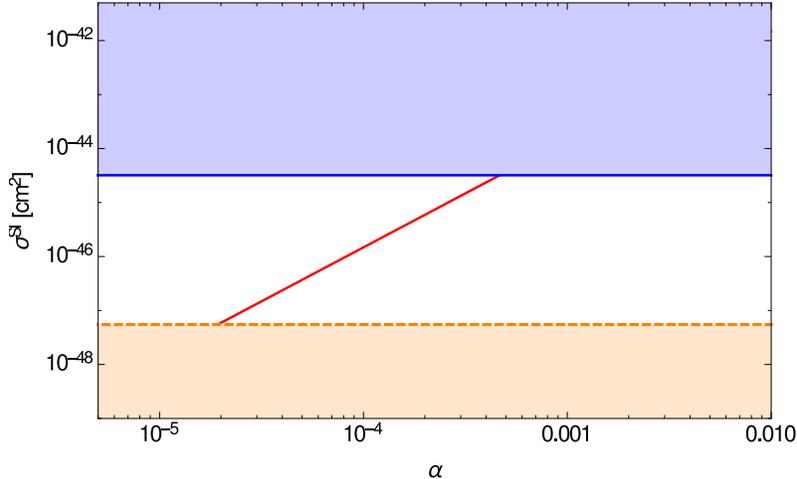} 
\end{center}
\caption{
The prediction of $\chi$-nucleon scattering cross-section as a function of
 the mixing angle of two Higgs bosons $\alpha$, for $m_{\chi} = 3$ TeV and $\alpha_{BL}=0.01$.
The blue shaded region is excluded by the most stringent bound from the PandaX-II experiment (2017)~\cite{PandaX-II}.
The orange shaded region represents the coherent neutrino background limit.
}
\label{fig:2}
\end{figure}

Although it is not stable, the RHN DM lifetime is extremely long and 
  its basic properties are the same as those of the Weakly Interacting Massive Particle DM. 
Hence, one may consider a possibility to directly and/or indirectly detect the DM in the standard methods. 
As far as direct detection is concerned,  
 in $\lambda_{\mathrm{mix}} \ll 1$ limit,
 the scattering cross-section with a nucleon through Higgs boson exchanges is negligibly small
 and beyond the reach of any planned experiment.
As is shown in Fig.~\ref{fig:2},
 the scattering cross-section with a nucleon through Higgs boson exchanges 
 is small enough to be consistent with the current bound for smaller Higgs mixing $\alpha \lesssim 10^{-3}$.  
For $\alpha \lesssim 10^{-5}$,
 the scattering cross section with a nucleon through Higgs boson exchanges is so small
 that signals from DM are buried under the coherent neutrino background.
A pair annihilation of the RHN DMs can be another source of cosmic rays. 
However, since this DM is a Majorana particle and its pair annihilation occurs 
  through the $s$-channel $Z^{\prime}$ boson exchange, 
  the annihilation cross-section today $(\sigma v_{\rm rel})_0$ is much smaller 
  than that in the early Universe $\langle\sigma v_{\rm rel}\rangle$ due to a very small small kinetic energy at present. 
Thus, the RHN DM annihilation cannot generate any detectable cosmic ray flux in the near future.

\section{Summary}

The DAMPE has just reported a measurement 
  of the flux of high-energy CREs in the energy range between $25$ GeV and $4.6$ TeV. 
Interestingly, the DAMPE data exhibit a peak excess of the CREs flux at an energy of around 1.4 TeV. 
For the DM interpretation of the DAMPE peak, we have considered 
  the minimal gauged $U(1)_{B-L}$ model with a RHN DM, 
  which is a minimal extension of the SM,
  to incorporate the neutrino mass matrix and a thermal dark matter.  
We have found that the origin of the DAMPE peak can be explained by the late-time decay of
 the RHN DM into $e^\pm W^\mp$ 
   with a DM lifetime $\tau_{\chi} \sim 10^{28}$ s and its mass $m_{\chi}=3$ TeV, 
   assuming a sub-halo near the Earth. 
This favored lifetime is almost same as the current bound on the lifetime from diffused gamma ray flux 
 measured by Fermi-LAT.
Hence, our model can be tested when the Fermi-LAT bound on the lifetime of decaying DM 
 becomes more stringent. 
The RHN DM communicates with the SM particles 
  through the $Z^\prime$ boson,  
  and its thermal relic abundance is controlled by only three free parameters: $m_{\chi}$, $\alpha_{BL}$, and $Z^\prime$. 
We have identified the parameter region to satisfy the constraints from the observed DM relic density, 
  the current LHC results and the perturbativity of $\alpha_{BL}$ up to the Planck scale, 
  which turns out to be very narrow. 
When we fix  $m_{\chi}=3$ TeV for the successful DM interpretation of the DAMPE peak, 
   we find the rest of the parameters to be $m_{Z^\prime} \simeq 6$ TeV 
   and  $0.00807\leq \alpha_{BL} \leq 0.0149$. 
As the $Z^\prime$ boson resonance search at the LHC continues,  
   the diagonal dotted line in Fig.~\ref{fig:1} will shift to the right 
   and our scenario will be tested in the future.


\section*{Acknowledgments}
The work of N.O. is supported in part by the United States Department of Energy (No.~DE-SC0013680).


\appendix




\end{document}